\documentstyle[prl,aps,floats,twocolumn]{revtex}
\newcommand{\Section}[1]{\section{#1}}
\begin{document}
\def\be{\begin{equation}}
\def\ee{\end{equation}}
\def\bea{\begin{eqnarray}}
\def\eea{\end{eqnarray}}
\def\E{{\rm e}}
\def\bearst{\begin{eqnarray*}}
\def\eearst{\end{eqnarray*}}
\def\peleven{\parbox{11cm}}
\def\peffec{\peight{\bearst\eearst}\hfill\peleven}
\def\pspace{\peight{\bearst\eearst}\hfill}
\def\ptwelve{\parbox{12cm}}
\def\peight{\parbox{8mm}}
\twocolumn[\hsize\textwidth\columnwidth\hsize\csname@twocolumnfalse\endcsname

\title
{Forced Burgers Turbulence in 3--Dimensions}

\author
{  Jahanshah Davoudi $^{c,d}$, A. Reza Rastegar$^{a,d}$ and
M. Reza Rahimi Tabar$^{b,c,d}$ }

\vspace{1cm}
\address
{\it $^a$ Department of Physics, Tarbiat-Moallem University, Tabriz, Iran
\\
$^b$ CNRS UMR 6529, Observatoire de la C$\hat o$te d'Azur,
BP 4229, 06304 Nice Cedex 4, France,\\
$^c$ Dept. of Physics , Iran  University of Science and Technology,
Narmak, Tehran 16844, Iran.
\\$^d$ Institue for Studies in Theoretical Physics and 
Mathematics
Tehran P.O.Box: 19395-5531, Iran,}  

%rahimi@netware2.ipm.ac.ir \\}

%\date{01/02/98}
\maketitle
%%%%%%%%%%%%%%%%%%%%%%%%%%%%%%%%%%%%%%%%%%%%%%%%%%%%%%
%ABSTRACT
%%%%%%%%%%%%%%%%%%%%%%%%%%%%%%%%%%%%%%%%%%%%%%%%%%%%%%

\begin{abstract}

We investigate non-perturbative results of inviscid forced Burgers equation 
supplemented to continuity equation in 
three--dimensions.
The exact two--point correlation function of density is calculated in 
three-dimensions. 
The two--point correlator $<\rho(\bf x_1) \rho(\bf x_2)>$ behaves as 
$ |{\bf {x_1 - x_2}}|^{-\alpha _3}$ and in the universal region 
$\alpha_3 = 7/2$ while in the non-universal region
$\alpha_3 = 3$. 
%In the universal region 
%we show that the angular dependence of the velocity correlation function
%satisfies the same equation, which is found in the instanton approach by
%Gurarie and Migdal {[Phys. Rev. E {\bf 54}, 4908 (1996)]}.
%The tail of the velocity increments is also found to behave as 
%$\frac{1}{r} \exp({-(\frac{|\Delta u|}{r})^3})$. 
In the non-universal region 
we drive a Kramers-Moyal equation governing the
evolution of the probability density function (PDF) of longitudinal 
velocity increments for three dimensional Burgers turbulence. In this region 
we prove Yakhot's conjecture {[Phys. Rev. E {\bf 57}, 1737 (1998)]}
for the equation of PDF for three dimensional Burgers turbulence.
We also derive the intermittency  exponents for  
the longitudinal structure functions and 
show that in the inertial regime one point $U_{rms}$ enters 
in the PDF of velocity difference. 
%Also we discuss the zero viscosity limit of Burgers turbulence and the  
% in mean field approximation and show that in this approximation 
% the viscous term in the vanishing viscosity limit do not contribute
% in the equation for the PDF's of velocity difference.

PACS numbers, 47.27.Gs and 47.40.Ki
\end{abstract}

%PACS numbers, 47.27.Gs and 47.40.Ki

\hspace{.2in}
%\newpage
]

\vspace {.5cm}
\Section{- Introduction}

%{\bf 1- Introduction}

Recently, tremendous activity has started on the 
non-perturbative understanding of turbulence 
%\cite{pol1,mig1,proc,mez1,pol2,fal,chert,
%gur,mez2,gaw1,ya2,gotoh,sinai2,gaw2,sinai1,sinai3
%bold1,bold2,rahimi,gross}.
[1-22].
A statistical theory of turbulence has been put
forward by Kolmogorov \cite{kol}, and further developed by others 
\cite{monin,dedo,ya1}.
The approach is to model turbulence using stochastic
partial differential equations.  
The simplest approach to turbulence is the Kolmogorov's dimensional  
analysis, which leads to the celebrated $ k^{- 5/3}$ law for the energy spectrum. 
This is obtained by decreeing that the energy spectrum depends neither on 
the wavenumber where most of the energy resides, nor on the wavenumber
of viscous dissipation. Kolmogorov conjectured that the scaling 
exponents are universal, independent of the statistics of large--scale
fluctuation and the mechanism of the viscous damping, when
the Reynolds number is sufficiently large. In fact the idea of universality
is based on the notion of the "inertial subrange". By inertial subrange
we mean that for very large values of the Reynolds number there is a wide 
separation between the scale energy input $L$ and the typical viscous
dissipation scale $\eta$ at which viscous friction become important and 
 the energy is turned into heat. 

In turbulent fluid flows,the physical quantities of interest, such as 
the velocity and density fields, display highly irregular fluctuations 
both in space and time. To study such processes, using a statistical approach 
is most natural. The statistics of any fluctuating quantity are described by 
its probability density function (PDF). One would, therefore, hope to be 
able to calculate the PDF's 
of the turbulent quantities directly from the 
equations of motion, i.e. for example to calculate the PDF of velocity 
fluctuations from Navier-Stokes equation. However this task is highly 
nontrivial and as of today has not been accomplished not even for the 
relatively simpler problems. The striking feature of developed turbulence
is its intermittent spatial and temporal behavior. The structures that arise
in a random flow manifest themselves as high peaks at random places and at
random times. The intervals between them are characterized by a low intensity
and a large size. Rare high peaks are responsible for PDF tails while the 
regions of low intensity contribute PDF near zero. 
Analytically the following properties manifest themselves in the
nonlinear bahavior of the scaling
exponents in the velocity difference structure functions (transvers and longtudinal) 
$S_n(r)$
i.e. $<(u(x+r) - u(x))^n>$. It is known that $S_n (r)$ behaves as 
$\sim r^{\zeta_n}$, where
$\zeta_n$ is a non-trivial function of $n$. This behaviour leads to stongly
non-gaussian (intermittent) probability distribution functions of
velocity difference \cite{uriel}. 
                                   
On the other hand Burgers equation, which describes the 
potential flow of the fluid without pressure, provides a wonderful laboratory
for testing new ideas and techniques in view of the study of fully
developed turbulence in the Navier--Stokes turbulence. These are two cases 
of non--linear stochastic equations which share in the same structure of the 
non--linearity. The important differences come from the nature of 
interaction (i.e. locality and non-locality) and the large scale structure. 
In the case of incompressible turbulence the interaction is non--local that is 
the very existence of the pressure causes the far regions of the flow to be coupled 
with together \cite{ya2} and leads to effective energy redistribution between 
components of the velocity field, while  
in the case of Burgers equation the pressure effects are absent. 
The non-linear term $({\bf v\cdot\nabla}){\bf v}$ 
in the Navier-Stokes and Burgers equations tends to form shocks and enhance 
intermittency while the additional pressure gradient term in N-S equation 
depresses intermittency. This senario has been given addressed by V. Yakhot 
\cite{ya2}. 
for constructing an approxiamte dynamical theory for N-S equation. 
In multi-dimensional Burgers turbulence the presence of large scale structures 
(shocks) forming a d-dimensional forth-like pattern is responsible for extreme
case of intermittency causing the saturation of the intermittency exponent to
$\zeta_n=1$. Large scale structures in "true" turbulence are similarly thought 
to be the origin of the experimentally observed intermittency, which is however 
much milder. The deep reason of this difference is thought to be probably
related to the "dimension" of the large scale singularities, which is $d-1$
in the d-dimensional Burgers turbulence and only $1$ for votex lines in 
hydrodynamical
turbulence.

The problem of forced and un-forcerd Burgers turbulence 
[28-30] has been  
attacked recently by various methods 
\cite{mig1,mez1,pol2,fal,gur,mez2,gaw1,ya2,gotoh,sinai2,gaw2,sinai1,sinai3,bold1,bold2,rahimi}.
The Burgers equation descibes a 
variety of non-linear wave phenomena arising in the theory of 
wave propagation, acoustics, plasma physics, surface growth 
, charge density waves, dynamics of the vortex 
lines in the high $T_{c}$ superconductor, dislocations in the disordered 
solids and formation of large--scale structures in the universe \cite{saichev,kardar,ver}. 
According to the recent theoretical \cite{mez1,pol2,ya2,bold1,bold2} and 
numerical works \cite{ya3}, it is known 
that the PDF for the velocity difference behaves differently 
in universal and non-universal regions. In the universal region i.e. 
the interval $|\Delta u| << U_{rms}$ and $r << L$, the PDF can be represented 
in the universal scaling form 
\be
P( \Delta u ,r) = \frac{1}{r^z} F(\frac{\Delta u}{r^z})
\ee
where $F(x)$ is a normalizable fuction and the exponent $z$ is related to the 
exponent of random force correlation $\eta$ as $z=\frac{\eta +1}{3}$.
In the region for $x=\frac{|\Delta u|}{r^z} >> 1$ the universal scaling 
function  $F(x)$ is given by the expression $F(x)\sim \exp(- \alpha x^3) $,
where $\alpha$ is some constant in one-dimension and it depends on the cosine
of angle between the vectors $ {\bf \Delta u}$ and $\bf r$ in the higher 
dimensions. On the other hands the PDF in the interval 
$|\Delta u| >> U_{rms}$ behaves as: 
\be
P(\Delta u ,r )= r \hskip .1 cm G(\frac{\Delta u}{U_{rms}})
\ee
which depends on the single-point $U_{rms}$ and therefore
is not a universal function.

In this paper we consider the inviscid forced Burgers equation in
3--dimensions
supplmented to continuity equation.
We find the exact two--point correlation functions of density in 
three-dimensions and show that $<\rho(\bf x_1) \rho(\bf x_2)>$ behaves as 
$ |{\bf {x_1 - x_2}}|^{-\alpha _3}$, where $\alpha_3 = 7/2$ in 
the universal region i.e. $|u(x) - u(x')| << U_{rms}$ where $U_{rms} $
is the $rms$ value of velocity fluctuations.  
%We show that the angular dependence of the velocity correlation functions
%satisfies the same equation, which is found in the instanton approach by
%Gurarie and Migdal \cite{gur} and find the tail of the PDF for velocity
%difference.
In the non-universal region i.e.  $|u(x) - u(x')| >> U_{rms}$
we drive the exact Kramers-Moyal equation governing on the
evolution of the probability density function (PDF) of longitudinal 
velocity increments for three dimensional Burgers turbulence. In this region 
we prove the Yakhot conjecture \cite{ya2} 
for equation of the PDF in multi-dimensional Burgers turbulence and prove 
the existence of the 
A-term and B-term in the equation for the PDF of longitudinal 
velocity difference. We also 
derive the intermittency  exponents for  
the longitudinal structure functions and show that how the 
$U_{rms}$ enters 
in the PDF of the velocity difference and breaks the Galilean invariance.
%We show that the Galilean invariance breaks spontaneously in the 
%non-universal region 
%and the $U_{rms}$ enter in the argument of PDF's in this region.
Furthermore we refer to the zero viscosity limit of 3D forced Burgers
turbulence. 
We claim that at least when the simplest terms satisfying the basic
symmetries 
of the problem 
are introduced for closing the disspiation term
the requirements of the homogeniety and isotropy would force the 
coeffiecient of
the proposed terms to vanish.  
So within this approximation the viscous term in the vanishing viscosity 
limit do not contribute in the equation for the PDF's of velocity difference
in the inertial range.

The paper is organized as follows:
In section 2 we derive the density-density correlators 
exponent in invicsid Burgers 
turbulence and the generating function for longitudinal velocity
difference 
in the universal part of the PDF .
%and show that the 
%angular dependence of PDF satisfies the same equation which has been found by 
%Gurarie and Migdal. 
In section 3 we 
determine the small scale statisics of longitudinal
velocity difference and derive the exact equation
for the PDF's of longitudinal velocity difference in the non-universal part of 
the PDF and derive 
the exact value of intermittency exponents 
in longtidinal structure functions. We show that the intermittency
problem in these systems can be investigated non-perturbatively and derive
the deformation of PDF's in length scales. It is shown that deformation of
PDF from large to small scales are completely described by Kramers-Moyal
equation.  

\vskip .3cm

\Section{-  PDF's of 3D Burgers Turbulence in the Universal Region } 

Our starting point is the 3D--Burgers  equation supplemented to 
continuity equation:

\be
{\bf u}_t + ({\bf u} \cdot \nabla ) {\bf u} = \nu \nabla^2 {\bf u}
+ {\bf f}({\bf x},t)
\ee
\be
 \rho_t + \partial_{\alpha} (\rho u_{\alpha})=0
\ee
for the Eulerian velocity $ {\bf u}({\bf x},t)$ and 
viscosity $\nu$ and density $\rho$, in 3--dimensions.
%Moreover at this level we intend to study the following equation 
%when viscosity contribution is absent .
The force $ {\bf f}({\bf x},t)$ is the external stirring force, which
injects energy into the system on a length scale $L$.
More specifically one can take, for instance a Gaussian distributed  
random force, which is identified by its two moments:

\bea
< f_\mu ({\bf{ x}},t)>& = & 0 \cr \nonumber \\
< f_\mu ({\bf{ x}},t)  f_\nu ({\bf x^{'}},t^{'})>& = & 
k(0) \delta (t-t^{'}) k_{\mu \nu}({\bf {x}- { x^{'}}})
\eea
where $\mu, \nu = x, y, z$. The correlation function 
$k_{\mu \nu}(r)$ is normalized to unity at the origin and decays 
rapidly enough where $r$ becomes larger or equal to integral scale $L$.
The quantity $k(0)$ measures the energy injected into the turbulent 
fluid per unit time and unit volume. $ {\bf f}({\bf x},t)$ provide also the 
energy flux in the $k-th$ shell as $\Pi_k = \Pi (r=k^{-1}) \simeq 
\int_{1/L} ^ k < | {\bf f}({\bf k})|^2> $, where $r$ belongs to the inertial 
range.

Eqs.(3) and (4) exhibit special type of nonlinear interactions.
It is hidden in the nonlinear term $ ({\bf u} \cdot \nabla ) {\bf u}$. 
The advective term couples any given scale of motion to 
all the large scales and the large scale contain most of the energy of the 
flows. This means that the large--scale fluctuation of turbulence production
in the energy--containing range couple to the small--scale dynamics 
of turbulence flow. In other words, the details of the 
large--scale turbulence production mechanism are important, leading 
to the non--univesality of probability distribution function (PDF) of 
velocity difference. However in the case that  $|u(x) - u(x')| << U_{rms}$
it is believed that the PDF for the velocity difference is not depended
to $U_{rms}$ and therefore the one-point  $U_{rms}$ does not appear in  
the velocity difference PDF. This region is known as
 the Galilean invariant (GI) region.
%In this section our aim is to find the generating function for velocity
%difference and hence the tail of PDF in the universal of GI- region.

Now let us consider the invisid 3D forced Burgers turbulence
that is we intend to neglect from the viscosity contribution in eq.(3).
The problem is to understand the statistical properties of velocity and
density fields which are the solutions of eq.(3) and (4). 
We consider the  following two--point
generating function:
\bea 
&& F_2 ({\bf \lambda_1, \lambda_2, x_1, x_2}) = \nonumber \\
 && <   \rho( \bf{x_1})  \rho(\bf{x_2})    \exp (
{\bf \lambda_1 .  u ( x_1) +  \lambda_2 .  u( x_2)} )>
\eea
where the symbol $< \cdots >$ means an average over various realizations of the 
random force. 
%Now one can show that $F_2$ satisfy the following equation, 

To derive an equation for $F_2$, we 
write the eq.(3) and (4) in two points $\bf x_1$ and $\bf x_2$ for 
$u_1,u_2, u_3$ and $\rho(x)$ 
and multiply the equations in $ \rho( { \bf{x_2}}) $, 
$\lambda_{1 x} \rho( { \bf{x_1}}) \rho( { \bf{x_2}}) $, 
$ \cdots $, $\lambda_{1 z} \rho( { \bf{x_1}}) \rho( { \bf{x_2}}) $ 
and $\rho( { \bf{x_1}})$, 
$\lambda_{2 x} \rho( { \bf{x_1}}) \rho( { \bf{x_2}})$, $\cdots$,
and $\lambda_{2 z} \rho( { \bf{x_1}}) \rho( { \bf{x_2}})$, respectively.
We add the equations and multiply the result by
$ \exp ({ \bf \lambda_1} . { \bf u (\bf x_1)} + {\bf \lambda_2 }. {\bf u({\bf x_2})} )$
and make average with respect to external random force, so we find: 

\bea
&& \partial _t F_2 + \sum_{\{i=1,2\} \mu= x,y,z}  \frac {\partial}{\partial \lambda_{i,\mu}}
\partial _{\mu_i } F_2 -
 \nonumber \\ && \sum_{\{i,j=1,2\} \mu,\nu = x,y,z} 
\lambda_{i, \mu} \lambda_{j, \nu} k_{\mu \nu} ({\bf x_i}- {\bf x_j}) F_2
= 0
\eea
%where $ D_2 $ is,
%\bea 
%D_2 & =& < \nu \rho( { \bf{x_1}}) \rho( { \bf{x_2}}) [ 
%{\bf {\lambda_1}} \cdot \nabla^2 {\bf{ u }}({\bf {x_1}}) + {\bf {\lambda_2}} 
%\cdot \nabla^2 {\bf {u}( {x_2})} ]  \cr \nonumber \\    
%&&\exp {( {\bf {\lambda_1}} \cdot {\bf{ u }}({\bf {x_1}}) + 
%{\bf {\lambda_2}} \cdot {\bf {u}( {x_2})} )}>
%\eea
where we have used the Novikov's theorem. 
The above equation is first driven 
by Polyakov \cite{pol2}. 
It is noted that the advection contributions are 
accurately accounted for in this equation.
%, it is not closed due to the 
%dissipation term i.e. $D_2$--term. Our aim now is to extend the assumptions 
%of the operator product expansion, Galilean and scaling invariance introduced 
%in [5,23] to take into account this term in $N$--dimensions. First we note 
%that the operator product expansions should be invariant under the rescaling
%of density i.e. $\rho \rightarrow \alpha \rho$. 
%Also the basic equations are 
%Galilean invariant. 
%On the other hand, final expression for $D_2$ must contain the  
%ultraviolet finite operators $ \nabla {\bf u}$, $\rho $
%and $ e^{\bf{\lambda \cdot u}}$.
%The only finite combination satisfying the rescaling  
%$\rho \rightarrow \alpha \rho$ is $\rho e^{{\bf {\lambda \cdot u}}}$ (see [23] for 
%more detail).  Theref0^R
%\be
%D_2 = a F_2
%\ee
%where $a$ is generally a function of $\lambda_1$ and $\lambda_2$.
Also we suppose that $k_{\mu \nu}$ has the following form:
\bea
k_{\mu \nu} ({\bf x_i} - {\bf x_j})& =& k(0) [1- 
\frac {| {\bf x_i} - {\bf x_j}|^2}{2 L^2} \delta_{\mu,\nu} \cr \nonumber \\
&-& \frac {({\bf x_i} - {\bf x_j})_\mu ( {\bf x_i} -  {\bf x_j})_\nu}{L^2} ]
\eea
with $k(0), L=1$. 
%Now, let us consider the eq.(5) in three dimensions. 
We change the variables as:
$\bf x_\pm = \bf x_1 \pm \bf x_2 $, $ \bf \lambda_+ = \bf \lambda_1 + \bf \lambda_2$ 
and 
$\bf \lambda_-  = \frac { {\bf \lambda_1} - {\bf \lambda_2} }{2}$ and 
and consider the spherical coordinates, so that 
$ x_- : (r,\theta, \varphi)$ and $ \lambda_-: (\mu, \theta ', \varphi')$.
Hence we find that the $ F_2 $ satisfies the following 
closed equation for homogeneous and isotropic Burgers turbulence: 
\bea
&&[ s \partial_r \partial_{\mu} - \frac{s(1-s^2)}{r \mu} \partial^2 _s 
+ \frac {1+s^2}{r \mu} \partial_s  
+ \frac {1-s^2}{\mu} \partial_r \partial_s \nonumber \\ &&  
+ \frac {1-s^2}{r} \partial_{\mu} \partial_s 
- r^2 \mu^2 (1 + 2 s^2)] F_2 =0
\eea
where  $ s = \cos{\gamma} =
 \cos{\theta} \cos {\theta'} + \sin {\theta} \sin{\theta'} \cos(\varphi - \varphi')$.

%Also we define the  variables
%$z$ and $s$, so that $ s=\cos{\gamma}$ and $z= r \mu$, therefore we find:  
%\be
%\partial_r = \mu \partial_z \hskip 1cm  \partial_\mu = r \partial_z 
%\hskip 1cm \partial_r \partial_\mu= z \partial_z ^2 + \partial_z
%\ee
%\bea
%&&\partial_\theta = [- \sin{\theta} \cos {\theta'} + \cos {\theta} \sin{\theta'} \cos(\varphi - \varphi')] \partial_s  
%\nonumber \\ &&
%\partial_{\theta'} = [- \sin{\theta'} \cos {\theta} + \cos {\theta'} \sin{\theta} \cos(\varphi - \varphi')] \partial_s  
%\nonumber \\  &&
%\partial_\varphi = - \sin{\theta} \sin{\theta'} \sin(\varphi - \varphi') \partial_s  
%\nonumber \\ &&
%\partial_{\varphi'} =  \sin{\theta} \sin{\theta'} \sin(\varphi - \varphi') \partial_s  
%\eea

%The $\mu$ dependence of the $a(\mu)$ anomaly must be chosen to conform the 
%scaling and can be changed depending on the scaling properties of the 
%force correlation functions. In general, in the case of isotropic 
%turbulence, stirring correlation
%function behaves as $k_{\mu \nu} \sim 1 - r^{\eta}$, where in our  
%case we have $\eta=2$. Therefore, $a$ must depend on $\mu$ as follows 
%$a(\mu) = a_0 {\mu}^{\sigma} $, where $\sigma = \frac {2 - \eta} {1+ \eta}$. 
%It is evident that for our case $a'$ is independent of $\mu$
%i.e. $a(\mu) = a_0$. The unknown parameter $a_0$ should be
%determined from the main requirements that the probability distribution 
%function, which is the Laplace transform of $ F_2$, be  positive, 
%finite and normalizable.
In homogeneous and isotropic Burgers turbulence, with stirring 
correlation as $ k(r) \sim 1- r^{\eta}$ (where in the our case 
i.e. eq.(8) we have $\eta=2$), we consider the universal scale--invariant
solution of eq.(9) in the following form:
\be
 F_2 (\mu , r) = g(r) F(\mu r^{\delta}) \hskip 1cm g(r) = r^{-\alpha_3}
\ee
Substituting the following form for the generating function fixes the
exponent $\delta$ as
$\delta = \frac {\eta + 1}{3}$ (In our case using eq.(8) we find $\delta = 1$).
Invoking to the scaling invariance of the inviscid Burgers equation and 
continuity equation, we assume the existence
of the density-density correlators with the
scaling form of introduced in (10). 
%where it can be found by
%the generating function in the limit of $\mu \rightarrow 0$.
$\alpha_3$ is
the exponent of two point correlation 
and the two point correlation of the density can be found by
the generating function in the limit of $\mu \rightarrow 0$.
%Indeed we suppose that the two--point correlation function of 
%density exists, and 
Therefore it is necessary to find such a solution for $F(\mu r^{\delta})$ 
which tends to a constant in the limit of $\mu\rightarrow 0$. 
%We will find sa solution for the generating function and that 
%garanties the existence of the density correlation for three dimensional case. 
Proceeding further we focus our attention to the longitudinal velocity
components,i.e. $s=1$, and hence accept the scaling ansatz 
$F(\mu r,s)=F(\mu r s)$.
The proposed form of the arguments will garantee that
$S_n(r,s)\sim s^n S_n(r)$ when $n<1$. Rewriting eq.(9) in terms
of the variable $z=\mu r s$, the following equation is obtained
in the limit of $s\rightarrow 1$:
\be
z \partial_z^2 F(z) + (3-\alpha_3) \partial_z F(z) - 3 z^2=0
\ee
It is interesting that the above equation was first derived 
by Polyakov \cite{pol2} for the problem of one dimensional Burgers
equation
in the {\it inviscid limit}.
In that work the effect of the viscose term is found in the limit
of $\nu\rightarrow 0$ and $r\ll L$ by appeal to the self-consistent
conjecture of operator product expansion. It is found that there are
only two terms generated by
the viscose term which are consistent with the symmetries of the
problem.These two anomaly terms will modify the master equation governing
over the generating function in a way such that a positive , finite
and renormalizable PDF can be found \cite{pol2}.  
A simple analogy between eq.(10) and Polyakov's result will reveal that
the coefficient of {\it b-anomaly} is replaced with the scaling exponent 
of density-density correlation through a simple linear relation.
In the problem of one-dimensional Burgers equation in the zero viscosity
limit the presence of the {\it b-anomaly} generated by viscosity term
is crucial for finding a positive PDF for velocity increments in the
universal regime and the requirement of positivity will fix the value of
anomaly coefficient \cite{pol2}. Boldyrev \cite{bold2} shows that one
could find a family
of solutions for different values of the b-anomaly coefficient if one
relaxed the homogeniety condition for the universal part of the PDF. The
value of this coefficient
is related to the algebraic decay of the left tail of PDF in the
universal regime.
Determination of that decay exponent has been a controversial subject for
which other     
methods have been developed and 
among them recent rigorous methods should be mentioned within which 
it is fixed to $7/2$ \cite{sinai1,sinai3}.
The interesting point is that our calculations in three dimensions show
that
when density fluctuations are taken into account there is no more any need
to the viscosity term for obtaining a positive PDF and even in the
inviscid problem it is possible to find a positive solution for the
longitudinal velocity increment PDF.
It is easy to show that the requirement of the positivity on the PDF will
fix the density-density scaling exponent to $\alpha_3=7/2$.
%Also we will derive the 
%exponent of the two--point correlation function of density by
%appeal to the requirement
%that the PDF should be positive, finite and normalizable. 
There are some strong constraints which would be considered for any
non-perturbative dynamical theory constructed for describing the 
intermittent
behaviour of the structure functions in the homogenious and isotropic 
turbulence flows. 
The most vital constraints are the homogeniety and isotropy which dictate 
the average of the velocity increment to be vanished.
Now it is straight forward to show $ < \bf {u}>_s = 0$, where 
$\bf{u}= \bf {u}(\bf{x}_1)- \bf {u}(\bf{x}_2)$ \cite{rahimi}. 

We have also extended our analysis to two--dimensional Burgers turbulence.
We define the variables as: 
$\bf x_\pm = \bf x_1 \pm \bf x_2 $, 
$ \bf \lambda_+ = \bf \lambda_1 + \bf \lambda_2$ 
and 
$\bf \lambda_-  = \frac { {\bf \lambda_1} - {\bf \lambda_2} }{2}$ and 
consider the polar coordinates, so that 
$ x_- : (r,\theta)$ and $ \lambda_-: (\mu, \varphi)$.
Now one can show that for isotropic and homogeneous Burgers turbulence,
two--point generating function $F_2$ satisfies the following 
equation: 

\bea
&&[ s \partial_r \partial_{\mu} - \frac{s(1-s^2)}{r \mu} \partial^2 _s 
+ \frac {s^2}{r \mu} \partial_s  
+ \frac {1-s^2}{\mu} \partial_r \partial_s \nonumber \\ &&  
+ \frac {1-s^2}{r} \partial_{\mu} \partial_s 
- r^2 \mu^2 (1 + 2 s^2) ] F_2 = 0
\eea
where $ s = \cos (\theta - \varphi)$. Similar to three--dimensions, 
we propose the scale invariant solution for $F_2$ as $F_2 = g(r) F(z,s)$
where $g(r)=r^{-\alpha_2}$ and $z= r \mu$. 
In the limit of $s=1$ once again it would be natural to
seek the scaling solutions as $F(r\mu,s)=F(r s\mu)$.
It will be straightforward to find that the eq.(11)
will again governs over $F_2(\mu r s)$ and the only difference 
originates from the coefficient of first derivative with respect to $z=\mu
r s$ which in this case is $2-\alpha_2$.
Using the positivity and 
normalizability condition of PDF we find 
$F(\mu rs) = \exp(z^\gamma )$ and show that  
$\alpha_2 = 5/2 $
and $\gamma=3/2$. 
%It is interesting that $f(s)$ in two-dimensions 
%also satisfies eq.(13) and instead of eq.(14), we get:
%\bea
%&&- s (1-s^2) f^{''}(s) + [(3+\alpha_2)- (2+\alpha_2) s^2] 
%f^{'}(s) \cr \nonumber \\
%&&+ \frac {3}{2} (\frac{3}{2} + \alpha_2 ) s f(s) = 0
%\eea
The exact values of the density-density exponents 
i.e. $ \alpha_2$ and $ \alpha_3$
are the main results of this part of our work which is 
derived self consistently.

Since the velocity difference PDF is 
the Laplace transformed
of $F$, one can readily deduce the right tail of the PDF as $ \frac{1}{rs}
\exp({- (\frac{\Delta u}{rs})^3)}$ 
(for $s=1$) in two and three dimensions
and in the limit 
$\frac{\Delta u}{r} \rightarrow + \infty $.
%, where $\alpha =\frac{2}{\sqrt{3}}$. 
This tail has been confirmed by several other approaches \cite{gur,mez2,sinai1}.
Left tail of the PDF is sensitive to the scaling exponent of the
density-density correlator and is given by
$\frac{1}{(\Delta u)^{(\alpha_3-1)}}$ when $\frac{\Delta u}{r}\rightarrow
-\infty$. 
At this stage we can not derive the PDF of density fluctuation.
The reason is that we have only two-point correlation functions of
the density
field while all of moments of density field are needed for this purpose.
In the one dimensional decaying Burgers turbulence 
it has been claimed that the density PDF would have 
some power law tail \cite{gotoh}
but this problem is open for the forced case.

\Section{-  PDF's of 3D Burgers Turbulence in Non--Universal Region }

In this section we consider the 3--dimensional Burgers turbulence 
in the non--universal region i.e. $|u(x) - u(x')| >> U_{rms}$,  
and derive the PDF of the longitdinal velocity difference 
and therefore the exponent of velocity structure functions.
As mentioned in the introduction recent works 
\cite{mez1,pol2,ya2,bold1,bold2} indicate that  
in the non-universal region the PDF of velocity difference
depends on the one-point $U_{rms}$ and therefore is not universal
which is meant to be sensitive on the details of large scale forcing.
This problem is known as the break down of Galilean invariance
in the non-universal region.
The force free Burgers equation is invariant under space--time translation,
parity and scaling transformation. Also it is invariant under Galilean
transformation, $x \rightarrow x + V t$ and $v \rightarrow v + V$, where
$V$ is the constant velocity of the moving frame.
Both boundary conditions and
forcing can violate some or all of symmetries of force free Burgers
equation.
However it is, usually assumed that in the high Reynolds number flow
all symmetries of the Burgers equation are restored in the limit
$r \rightarrow 0$ and $ r >> \eta$, where $\eta$ is the dissipation
scale where the viscous effects become important. This means that in 
this
limit the root--mean square velocity fluctuations $U_{rms}=
\sqrt{<v^2>}$
which is not invariant under the constant shift $V$, cannot enter the 
relations
describing moments of velocity difference. Therefore the effective
equations for the inertial--range velocity correlation functions must 
have
the symmetries of the original Burgers equations. For many years this
assumption was the basis of turbulence theories. But based on the recent
understanding of turbulence, some of the constraints on
the allowed turbulence theories can be relaxed \cite{pol2,ya2,bold1,bold2}.

%The force free Burgers equation is invariant under Galilean
%transformation, $x \rightarrow x + V t$ and $v \rightarrow v + V$, where
%$V$ is the constant velocity of the moving frame.
%Both boundary conditions and
%forcing can violate this symmetry of force free Burgers
%equation.
%However it is, usually assumed that in the high Reynolds number flow 
%(or zero viscosity limit)
%this symmetry of the Burgers equation are restored in the limit
%$r \rightarrow 0$ and $ r >> \eta$, where $\eta$ is the dissipation
%scale where the viscous effects become important. This means that in 
%this
%limit the root--mean square velocity fluctuations $u_{rms}=\sqrt{<v^2>}$
%which is not invariant under the constant shift $V$, cannot enter the
%relations describing the moments of velocity difference. 
%Therefore the effective equations for the inertial--range velocity 
%correlation functions must have
%the symmetries of the original Burgers equations. For many years this
%assumption was the basis of any turbulence theories. But based on the recent
%understanding of turbulence, some of the constraints on
%the allowed turbulence theories can be relaxed [1].
%The mstlgO@
%in not known so far.

According to recent Yakhot modeling of Burgers and N--S turbulence
this symmetry breaks in a hard way i.e. the $U_{rms}$ is entered
explicitly in 
the equation of PDF for velocity difference.
In the following we aim to show how this symmetry breaks 
in the sense that the one-point $U_{rms}$ enters in the argument of 
the PDF in non-universal region.
Also since we are interested in the scaling  of the
longitudinal structure function $S_q = < (u(x+r) - u(x))^q> \equiv < u^q >$,
where $u(x)$ is the $x$-component of the three-dimensional velocity
field
and $r$ is the displacement in the direction of the $x$-axis and the
probability density $P(u,r)$ for homogeneous and isotropic turbulence.
In the non--uiversal region using the eq.(11) one can observe that the PDF
for the velocity difference in the invicsid Burgers turbulence in two 
and three dimensions 
satisfy the following closed equation:
\bea
 && \frac{\alpha_d s}{r} \partial_u u P - s
\partial_u u \partial_r P - \frac {s (1-s^2)}{r} \partial_s ^2 P \cr \nonumber \\
& +&  \frac {d-2+s^2}{r} \partial_s P
-\frac{\alpha_d}{r} (1-s^2) \partial_s P +
(1-s^2) \partial_s \partial_r P   \cr \nonumber \\
&-&\frac {(1-s^2)}{r}\partial_u u\partial_s P
 +  r^2(1+2s^2) \partial_u ^3 P=0
\eea
where $ s$ equal to $ \cos (\theta - \varphi)$ in two-dimensions
and $ s= \cos{\gamma} =
 \cos{\theta} \cos {\theta'} + \sin {\theta} \sin{\theta'} \cos(\varphi - \varphi')$
 in three-dimensions. 
For determining the statistics of the small-scale
 and the behaviour of PDF 
 in the inertial range, i.e. $\eta<< r << L$,
 %, where $\eta$ is the dissipation 
 %length and $L$ is the integral length, 
 we can ignore the forcing term 
 in eq.(19). The reason for neglecting the force term is that 
the forcing contribution is 
 in the order of $ r^{2}/L^{2} $ so in the inertial range we can 
 safely drop the corresponding term. However we need to consider the effect
 of forcing by matching the PDF in the inertial range with the
 PDF in the integral scale.  
Since we are interested in equation of the 
 longitudinal PDF we have to consider the above 
 equation in  the limit of $s \rightarrow 1$.
This limiting is not trivial and needs to be considered more carefully.
 The contributions of different
terms of
the PDF equation in the limiting when $s \rightarrow 1$
is determined by the corresponding equation of the structure function. 

%Then the limiting would clearly   
%show the role of different terms in the structure function equation. 
%For solving 
%that equation one can seperate the angular part from which the power law
%solution arises. 
%Now substituting the angular parts in the original equation 
%will determine the role of different terms in the limit of $s\rightarrow 1$.
%It will be cleared out that different terms are playing the roles of source and sink
%in the original equation where naive treatment of the limitation would make to
%loose some terms which are very important in getting the correct result for 
%the structure function.
Assuming that the all of the moments of velocity
difference exist, the structure 
functions $S_n$ for given angle $\gamma$ ( or $s=\cos ({ \gamma})$)
satisfies the following  
closed equation:
\bea
&&[s n + (1-s^2) \partial_s ]  r  \partial_r S_n - n s \alpha_d S_n 
 \nonumber \\ &&
- s(1-s^2) \partial_s ^2 S_n + (d-2+s^2) \partial_s S_n 
 \nonumber \\&&
- \alpha_d (1-s^2) \partial_s S_n 
+ n (1-s^2) \partial_s S_n  \nonumber \\&&
+ r^3 n (n-1) (n-2) (1+2s^2) S_{n-3}
= 0
\eea
The forcing contribution to the above equation is the last term i.e. 
$r^3 n (n-1) (n-2) (1+2s^2) S_{n-3} $ and this term has not any 
contribution in the exponent of structure function. However the amplitude 
of the structure functions do depend on the details of forcing. 
We mean that 
the scaling ( multifractal ) exponent of the structure function is not
related to the forcing term and is related to the 
structure of non-linearity and the transverse contributions 
in the Burgers equation.
For solving 
the eq.(20) one can seperate the angular and scale dependent parts 
 of $S_n(r,s)$. The calculations give rise to the result that
 for $s\rightarrow 1$
the structure functions have the following form 
\be
S_n(r,s)\rightarrow s^n S_n(r),
\ee
where $S_n(r)=<(u(x+r)-u(x))^n> \equiv < u^n >$.
Factorizing the angle and scale
dependences is known also for the N--S turbulence too \cite{ya2,monin}. 
The proposed form for structure functions dictates that
in the limit when $s \rightarrow 1$ the probability distribution of velocity 
increments satisfies the scaling form, 
$P(r,u,s) \rightarrow \frac{1}{s}P(r,u/s)$ in a sufficient way.
It is clear that when $ s \rightarrow 1$, $P(r,u)=P(-r,-u)$ and it 
satisfies the following equation in d=2 and 3-dimensions.
\be
[- \frac {\partial}{\partial u} u - B] \frac{\partial}{\partial r} P +
\frac{A}{r} \frac {\partial}{\partial u} u P + 3 r^2 \frac {\partial^3}{\partial u^3} P
=0
%\frac{(\alpha_d - (d-1))}{r} \frac {\partial}{\partial U} U P=0
\ee
where $P$ is the longtudinal velocity difference $PDF$, 
and $B$ approaches to zero as $O(1-s^2)$ and giving address to $\zeta_3=1$
would give A=1.
The above equation is correct for three and two dimensions and 
it does not depend on the dimensionality. 
This form of equation for the PDF has been 
conjectured recently by Yakkot for the multi-dimensional Burgers turbulence \cite{ya2}.
It is noted that the forcing contribution in the above equation
is $ 3 r^2 \partial_{u}^3 P$ and it is irrelevant in the small scale 
$r \rightarrow 0$. We will deal with the forcing contribution
as a boundary condition of PDF in the large scales and this term 
is responsible for the breakdown of the Galilean invariance .
%and will enter
%the $U_{rms}$.
Also in the non-universal region we find the density-density exponent 
to be $\alpha_d = d$ in the two and three-dimensions.
Now it is easy to see that the eq.(8) can be written as
a scale--ordered exponential:
\begin{center}
$P(u,r)={\cal T}(e_{+}^{\int_{r_0}^{r} dr' L_{KM}(u,r')} P(u,r_0))$
\end{center}
where $L_{KM}$ can be obtained by computing the inverse operator \cite{jhd}.
Using the properties of scale--ordered exponentials the conditional 
probability density will satisfy the Chapman-Kolmogorov
equation. Equivalently we derive that the probability density
and as a result the conditional probability density of longitudinal
velocity increments satisfies a Kramers-Moyal evolution equation:
\be
-\frac{\partial{P}}{\partial{r}}=
\sum_{n=1}^{\infty} {(-1)}^{n}\frac{\partial^{n}}{\partial{u}^{n}}(D^{(n)}(r,u)P)
\ee
Where $D^{(n)}(r,u)=\frac{\sigma_{n}}{r} u^{n}$ \cite{jhd}.
We find that the coefficients $\sigma_{n}$ depend on $A$ and
$B$ through the relation
$\sigma_{n}=(-1)^{n}\frac{A}{(B+1)(B+2)(B+2)...(B+n)}$. 
%which are given by the 
%recursion relations
The same equation (i.e. eq.(23)) obviously governs over the 
conditional PDF too but with another boundary condition, i.e. 
$P(u,r | u',r)=\delta(u-u')$. 
%Since the information of the joint PDF
%deformation in scale is contained in the two-point conditional PDF
%in a Markovian process in scale then just by knowing the functional form of 
%the $KM$ operator, every information about the joint PDF's in the 
%inertial range can be dealt with  in an easy way. 
For a simple case we proceed to find velocity difference PDF $P(u,r)$ 
just by using the same 
line of reasoning and the well known Bayesian rule. So
\be
P(u,r) = \int P(u,r|u',L) P(u',L) du'
\ee
where $P(u',L)$ is assumed a Gaussian in the integral scale. Since 
\be
P(U,r| u',r)={\cal T}(e_{+}^{\int_{r_0}^{r} dr' L_{KM}(U,r')} \delta(u-u'))
\ee

The proposed $KM$ operator has an important property,i.e.
\be
L^+ _{KM} u^m = \zeta_m u^m
\ee
where $\zeta_m$ is the scaling exponent of the longitudinal 
velocity difference $S_m$.
Plugging the scale ordered form of the conditional PDF in the eq.(24), we 
will get:
\be
P(u,r) =(\frac{r}{L})^{ L^+ _{KM} (u)} P(u,L)
\ee
Expanding the assumed gaussian form of $P(u,L)$ in terms of $u$ and using (26)
ends with,
\bea
P(u,r) & =& \sum _{m=0} ^{\infty} \frac{\exp[\ln(\frac{r}{L}) \zeta_{2m}] }{m !}
(\frac {u}{U_{rms}})^{2m} (-1)^m \cr \nonumber
&=& {\frac{r}{L} }^{\zeta_{2m}}
\sum _{m=0} ^{\infty} \frac{ (-1)^m  }{m !} (\frac {u}{U_{rms}})^{2m}
\cr \nonumber &=&
(\frac{r}{L})^{\zeta_{2m}} exp(- (\frac {u}{U_{rms}})^{2}) 
\eea
where $\zeta_{j} =1$. We should stress that the same structure is tractable 
for the other PDF's in the integral scale other than the simple Gaussian form.
This result could be derived from eq.(22) by direct calculations even
without consulting the KM form of the evolution operator too.
This form verifies the proposed form of the PDF 
in the inner scales where $\eta << r << L$ and
resonates with numerical observation in the one dimensional
Burgers turbulence \cite{ya3}, where the non-universal part of the PDF
fits with $
P(\Delta u ,r )= r \hskip .1cm G(\frac{\Delta u}{U_{rms}})$. The same results 
was derived recently with additional assumption by Yakhot \cite{ya2}.
An interesting point with respect to the possible GI breaking mechanisms
should be refered. In our calculation we have shown that $U_{rms}$ 
and the one point information has revealed itself because of the matching
between the inertial range velocity increment PDF and the integral scale 
velocity increment PDF. Because the  
the variance of the velocity increments PDF in the integral scales and for 
the large $L$ is in the order of the variance of the {\it one point} PDF 
we observe
such a breakdown mechanism even with the lack of the forcing contributions 
in the $L_{KM}$. 
This way of breaking which we would like to call as soft 
GI breaking can be accompanied with the effects of the forcing where $U_{rms}$
will be entered ($U_{rms}=(k(0)L)^{1/3}$) in the equation of the PDF itself.
So the second way of the GI breaking which may be refered to as a hard 
GI breaking
is linked to the explicit dependence of $L_{KM}$ to $k(0)$. These parts of
the 
Kramers Moyal operator causes some scale dependence for the Kramers-Moyal
coefficients and the explicit calculation of the velocity increment PDF
in terms of the conditional ones is not a trivial task. 
The intricate part of the manipulations is related to the non-commutativity
of the GI breaking parts of the evolution operator with the GI invariant
parts so that the scale-ordered exponential can not be simply operated
on the integral scale PDF.
However it has been 
shown that these parts would cause non-universal behaviours of the amplitudes
in the velocity increment structure functions. That is for $r \rightarrow 0$,
\be
S_n(r)=A_n r^{\zeta_n}+3A_{n-3}\frac{n(n-1)(n-2)}{n+B}
\frac{r^{3+\zeta_{n-3}}}{3+\zeta_{n-3}-\zeta_{n}}
\ee
where $\zeta_{n}=\frac{An}{(n+B)}=1$.
This leads to
the non-universality of the PDF shapes in the inertial range \cite{ya2}.
In general both of these mechanisms may be relevant for the {\it scaling} 
of the inertial range PDF with $\Delta u/u_{rms}$ 
but their clear roles would be determined by explicit calculations.

The $A$ coefficient in eq.(22) is responsible  for the scaling of the structure functions
while the $B$ term is an infinitesimal coefficient which its value is
responsible for
$n$ independence of the scaling exponents.
One should note that according to our calculations we have derived the following
terms just by writing the whole equation which is governed over the PDF and 
then taking all the source terms which are proportional to the derivatives
of 
the PDF with respect to the angle $s$. So it is the most important result of 
our
calculations which resembles that without consulting to the conjectures
for 
introducing the scaling terms in 
the PDF equation 
%as was suggested recently by Victor Yakhot \cite{ya} for describing
%the intermittency of the structure functions of longitudinal velocity increments 
%in fully developed turbulence,
all the conjectured terms \cite{ya2}
could be driven just by carefully writing the $s$ dependence of PDF 
and then taking the 
limit $s\rightarrow 1$. So the $A$ term comes out to be $\alpha_d-(d-1) = 1$ and $B$ term
approaches to zero as $1-s^{2}$. These are in complete coincidence with the 
proposed values which was derived in Yakhot theory for the Burgers turbulence \cite{ya2}.
Using the above equations one can show that the $S_n(r)$ scales with $r$
as $r^{\zeta_n}$ where $\zeta_n =\alpha_d - (d-1)=1$.

According to Eq.(16) for the universal part in the three dimensions we
have found 
$\alpha_3=7/2$ while in the nonuniversal part we have got
$\alpha_3=3$.
Comparing the value of the density-density exponent $\alpha_d$ in 
universal and non-universal regions shows a small deviation in the two
parts.  
The dependence of the density density exponent to the specific parts of the
velocity increment PDF would be revealed if one takes a closer look at 
the original generating function. Considering the eq.(10), $g(r)$ is
referring to
density-density correlation function conditioned on a fixed
value of 
the velocity increment. So it's scaling exponent should depend on the
behaviour of the
velocity increment PDF too. 
%It is interesting that when one considers the 
%higher dimensions the difference will decrease until a critical dimension 
%$d_c=2.84307...$ is 
%reached where  
%the two exponent will coincide. Above the critical dimension this
%seperation 
%of the
%exponents will start again and they will differ as the dimension grows
%up.
%This picture might be related to the previous calculations of
%d-dimensional 
%turbulence \cite{uriel3,naji}.
%Neglecting the forcing term the scaling exponent
%of $S_3$ has not any deviation from exact results given by the equation 
%itself and this gives the $\alpha_3 \simeq 3$ and
%$\alpha_2 \simeq 2$, saying in another way, using the.

\Section{- Discussion } 
The first point which we wish to address is the role of viscosity
when $\nu \rightarrow 0$ instead of the inviscid limit.
It is well known that the viscosity contributions makes the 
generating function equation to become unclosed .
To our knowledge the only method for dealing with the viscosity
contribution is the groundbreaking method of Polyakov \cite{pol2} in the 
one dimensional forced Burgers turbulence.
In this method the overall contribution of the viscosity is assumed to
generate some other terms consistent with the symmetries of the original
equations and their presense just renormalises the terms 
contributed by advective derivative in the equation of generating function. 
%It has been stressed that without introducing the generated terms a positive,
%finite and normalisable PDF would not be developed \cite{ya2}.
Boldyrev \cite{bold2}  
generalised the Polyakov's method for the Burgers turbulence 
supplemented with continuity equation in one dimension by 
invoking to the same
strategy from which behaviour of the PDF tails in one 
dimension was given.
%However he neither could find the scaling exponents of the structure functions
%nor could find the non universal parts of the velocity increment PDF.
Keeping the viscous term in the three dimensional Burgers equation and 
with continuity equation we have examined the same closure for the 
viscous contribution.
The only relevant term consistent with the symmetries 
developed as a generalisation of the
Boldyrev work \cite{bold2} is $D_2=aF_2$ 
(see \cite{bold2,rahimi} for more detail).
However one can prove that in this framework   
the strong constraint of homogeniety, i.e. 
$<\bf{u}> = <\bf {u}(\bf{x}_1)- \bf {u}(\bf{x}_2)> = 0$
will fix the vlaue $a=0$ \cite{naji}. 
So at least this closure has no contribution in PDF equation and
the results will not be different from the inviscid calculations.

The second problem which we want to discuss about is related to the 
density probability function in general. As far as we know
there is one simulation being done by Gotoh and Kraichnan \cite{gotoh}
for finding the tail of the density PDF in one dimension and 
for the decaying burgers turbulence .
For the high density regime they have found a power law tail for the 
density PDF. It is generally accepted that in one dimension the mass 
accumulates in the shock regimes and the shock statistics will determine 
the density profile in the stationary regime. However we are not aware of  
such simulations for the forced problem even in one dimension.  
In higher dimensions because the nature of the singularities in the velocity flow 
are more involved \cite{saichev} the simple picture in one dimension regarding the mass accumulation 
in the singularities of the velocity profile can not be conducted in a trivial way.
Apart from the density PDF there is not any simulation for investigating any multi-point
corralation function of the density in higher dimensions.
For $d=1$ Boldyrev \cite{bold2}
has reported about a simulation on which the exponent of 
the two point correlation function of the density has 
predicted to be $\sim 2$.
Any attempt for simulating the forced Burgers equation with the continuity equation
would be valuable for clearing out the outcomes of our paper .
%We should stress that in our calculations the existence of the density 
%correlations have been assumed a priori however 

%In this paper we have given a descriptive way for explaining the 
%intermittency of the 3 dimensional forced Burgers turbulence.
%longitudinal structure functions by interrelating the deformation of the 
%PDF with the Kramers-Moyal evolution of the PDF in scale.
%In the Yakhot theory for incompressible fully developed turbulence 
In this paper we have proved the Yakhot conjecture about the PDF equation 
for the multi-dimensional Burgers turbulence. However 
he has 
shown that  
%modeling of such flows, 
considering $B\sim 20$ and using the exact result $\zeta_3=1$ as a boundary 
condition over final PDF, the 
incompressible fully developed turbulence can be modeled \cite{ya2}. 
From his theory he finds the multiscaling
of fully developed turbulence compatible with the experimental and simulation
results. In the centrepiece of that work it is emphasised that the effect 
of pressure just
renormalises the values of $B$ and creates the $A$ term which bahave as 
source and sink. According to our results we have shown that the $A$ term 
could be 
generated just by considering the angular dependences of PDF and even 
without pressure finite scaling exponents are derived and the terms responsible 
for scaling are present in the equation of the stationary PDF. 
It would be illuminating to seek whether, 
entering the effect of pressure just renormalises the values of $A$ and $B$
coefficients without adding any new terms in the PDF equation.
We beleive that the effect of pressure in the velocity 
intermittency can be followed if all the informations inherent in the 
correlations
of different moments of density field are given .
Other important perspective which we are pursuing is related to the 
deformation of transvers structure functions and a clear picture
about the transition between the longitudinal to transverse PDF.
Consequently this would give a thorough underestanding about the issue of
intermittency in the transverse structure functions \cite{ya2d}.

Our exact results are also appliciable to the description 
of fractal--nature of Interstellar Medium (ISM) if we accept 
that the Burgers equation supplemented with continuity eqaution is
a good candidate for modelling it \cite{ver}. 
In the framework of the Burgers--turbulence theory of interstellar 
medium we derive the scaling relation $M(R) \sim R^{d_H}$ for the mass on a 
region of size $R$, and the value of the $d_H$ can be predicted in this framwork 
which is $d_H=3/2$ .
The possible relation 
of ISM and Burgers turbulence will be disscused elsewhere \cite{rahimi3}.

We would like to thank 
Uriel Frisch, Mehran Kardar and Victor Yakhot for useful discussions 
and S. Rouhani and D.D. Tskhakaya for important comments.
M.R. Rahimi Tabar wish to acknowledge 
the support of the French Ministry of Education for his visit
to the Observatoire Cote d'Azur and 
U. Frisch for his kind hospitality.  

\vskip -.5cm
%\newpage 

\end{document}